\documentclass[aps,pre,twocolumn]{revtex4-1}  

 \usepackage{graphicx}
 \usepackage{amsmath,amssymb}
 \usepackage{bm}
 \usepackage{tcolorbox}
 \usepackage{hyperref}
 \usepackage{subfigure}

\def\bea{\begin{eqnarray}}
\def\eea{\end{eqnarray}}
\def\tg{\tilde{\gamma}}
\def\tb{\tilde{\beta}}

\def\nn{\nonumber}

\def\f{\frac}

\definecolor{dgreen}{rgb}{0,0.5,0}

\begin{document}

\title{COVID-19: Analytic results for a modified SEIR model and comparison   of different intervention strategies}

\author{Arghya Das, Abhishek Dhar, Srashti Goyal, Anupam Kundu, Saurav Pandey}
\affiliation{International Center for Theoretical Sciences, Tata Institute of Fundamental Research, Bangalore-560089, India}

\begin{abstract}
The Susceptible-Exposed-Infected-Recovered (SEIR) epidemiological model is one of the standard models of disease spreading. Here  we analyse an extended SEIR model that accounts for asymptomatic carriers, believed to play an important role in COVID-19 transmission. For this model we derive a number of analytic results for important quantities such as the peak number of infections, the time taken to reach the peak and the  size of the final affected population. 
 We also propose an accurate way of specifying initial conditions for the numerics (from insufficient data) using the fact that the early time exponential growth is well-described by the dominant eigenvector of the linearized equations.
Secondly we explore the effect of different intervention strategies such as social distancing (SD) and testing-quarantining (TQ). The two intervention strategies (SD and TQ) try to reduce the disease reproductive number, $R_0$, to a target
value $R^{\rm target}_0 < 1$, but in distinct ways, which we implement in our model equations. We find that for the same  $R^{\rm target}_0 < 1$, TQ is more efficient in controlling the pandemic than SD. However, for TQ to be effective, it has to be based on contact tracing and our study quantifies the required ratio of tests-per-day to the number of new
cases-per-day.  Our analysis shows that the largest eigenvalue of the linearised dynamics provides a simple understanding of the disease progression,
both pre- and post- intervention, and explains observed data for many countries.  We apply our results  to the COVID data for India to obtain heuristic projections for the course of the pandemic, and note that the predictions strongly depend on the assumed fraction of asymptomatic carriers.
\end{abstract}


\maketitle

The COVID-19 pandemic, that started in Wuhan (China) around December 2019 \cite{Huang2020, WHO2020},  has now affected almost every country in the world. The total number of confirmed cases on July 30, 2020 were close to $17.5$ million with close to $686,000$ deaths \cite{Worldometer}. One of the serious concerns presently  is that there is as yet no clear picture or consensus  on the future evolution of the pandemic. It is also not clear as to what is the ideal intervention strategy that a government should implement, while also taking into account the economic and social factors. The role of mathematical models has been to   
 provide guidance for policy makers \cite{Kucharski2020,prem2020,ferguson2020,Tang2020,colaneri2020,gatto2020,india1,india2,india3,india4,india5,frank2020a,piovella2020,israel2020,nigel2020}.

One of the standard epidemiological model is the SEIR model~\cite{li2018} which has four compartments of Susceptible ($S$), Exposed ($E$), Infected ($I$) and Recovered ($R$) individuals with $S+E+I+R=N$ being the total population of a region (the model can be applied at the level of a country or a state or a city and is expected to work better for well-mixed populations).  The SEIR model is parameterized by the three parameters  $\beta$,~$\sigma$   and $\gamma$ that specify the rates of transitions from $S\to E$, $E \to I$ and $I \to R$  respectively. In terms of the data that is typically measured and reported, $R$ corresponds to the total number of cases till the present date, while $\gamma I$ would be the number of new cases per day. The number of deaths across different countries is  some fraction ($\approx 1-10 \%$) of $R$ \cite{data2} while the number of hospital beds required at any time would be $\approx {\rm new~cases~per~day} \times {\rm typical~days~to~recovery}$.  An important parameter characterizing the disease growth is the reproductive number $R_0$ \cite{Giesecke, Andersson} --- when this has a value greater than $1$, the disease grows exponentially. Typical values reported in the literature for  COVID-19 are in the range $R_0=2-7$~\cite{reviewR0}.  For the  SEIR model one has $R_0=\beta/\gamma$ \cite{Andersson}.

The two main intervention schemes for controlling the pandemic are social distancing (SD) and  testing-quarantining (TQ). Lockdowns (LD) impose social distancing and effectively reduce contacts between the susceptible and infected populations, while testing-quarantining means that  there is an extra channel to remove people from the infectious population.  These two intervention schemes have to be incorporated in the model in distinctive ways \cite{Tang2020,colaneri2020} --- SD effectively changes the  infectivity  parameter $\beta$ while TQ changes the parameter $\gamma$.  Intervention schemes attempt to reduce $R_0$ to a value less than $1$. In the context of the SEIR model with $R_0=\beta/\gamma$,  it is clear that we can reduce it by either decreasing $\beta$ or by increasing $\gamma$. In this work we point out that for the same reduction in $R_0$ value, the effect on disease progression can be quite different for the two intervention strategies. 

Here we analyze an extended  version of the SEIR model which incorporates the fact that  asymptomatic or mildly symptomatic individuals \cite{Leung2018,Tang2020,colaneri2020,gatto2020} are believed to play a significant role in the transmission of  COVID-19. 
Our extended model considers eight compartments of Susceptible ($S$), Exposed ($E$), asymptomatic Infected ($I_a$), presymptomatic Infected ($I_p$), and a further four  compartments ($U_a,D_a,U_p,D_p$), two  corresponding to  each of the two infectious compartments. These last four classes comprise of individuals who have either recovered (at home or in a hospital) or are still under treatment or have died --- they do not contribute to spreading the infection.
We do not include separate compartments for the number of hospitalized and dead since these extra  details would not affect our main conclusions.  

For this extended SEIR model we first provide,  for the case with $R_0 > 1$,  analytic expressions for peak infection numbers, time to reach peak values, and asymptotic values of total affected populations. These would provide useful guidance on disease progression. The linearized dynamics is accurate when the total affected population is small. We  propose an accurate way of specifying initial conditions for the numerics (from insufficient data) using the fact that the early time exponential growth is well-described by the dominant eigenvector of the linearized equations.
We next discuss the performance of  two different intervention strategies (namely SD and TQ) in the disease dynamics and control. The aim of interventions is to reduce the reproductive number from it's free value $R_0$ to a target value $R_0^{\rm target}<R_0$.
We study  both the cases, of  strong interventions ($R_0^{\rm target}<1$ aimed at disease suppression,) and that of weak interventions ($R_0^{\rm target} \gtrsim 1$, aimed at disease mitigation). 
Apart from the reproductive number, $R_0$, an important parameter is the largest eigenvalue of the linear dynamics, which we denote as $\mu$.  For $R_0^{\rm target}  > 1$, we have $\mu >0$ and this gives us the exponential growth rate (doubling time $\approx 0.7/\mu$). On the other hand, for  $R_0^{\rm target}  <1$, the corresponding $\mu$ is less than $0$ and this tells us that infections will decrease exponentially. We find that, for the same reduction of $R_0^{\rm target}$ to a value less than $1$, the corresponding $\mu$ magnitude can be very different for different intervention schemes.  A larger magnitude of $\mu$, corresponding to a faster suppression of the pandemic, is obtained from TQ than that from SD. 
An important question for disease control is as to how much testing is required. In our work  we relate the parameters of the model related to  TQ to  testing rates and point out that for TQ to be successful: (a) it has to be based on contact-tracing and (b) it is necessary that testing numbers are scaled up according to the number of new detected cases. Finally, we show that many of our  results  provide a qualitative  understanding of COVID-19 data from several countries which have either achieved disease suppression or mitigation.

The rest of the paper is structured as follows. In Sec.~\eqref{sec:model} we define the extended SEIR model. 
A number of analytic results for the linearized model as well as the full nonlinear system are presented in Sec.~\eqref{sec:analytics}. 
In Sec.~\eqref{sec:interventions}, we discuss in detail how intervention strategies such as social distancing and testing-quarantining can  be incorporated into the model, we comment on how real testing numbers enter the  model parameters. We also present numerical studies comparing different intervention protocols and make  qualitative comparisons of  the predictions of the SEIR model with real data on the COVID-19 pandemic. 
In Sec.~\eqref{sec:india} we make some heuristic predictions in the Indian context.  We summarize our results in Sec.~\eqref{sec:conclusions}.

\section{Definition of the extended SEIR model}
\label{sec:model}

\begin{figure}[t]
\center
\includegraphics[width=8.3cm,height=5.5cm]{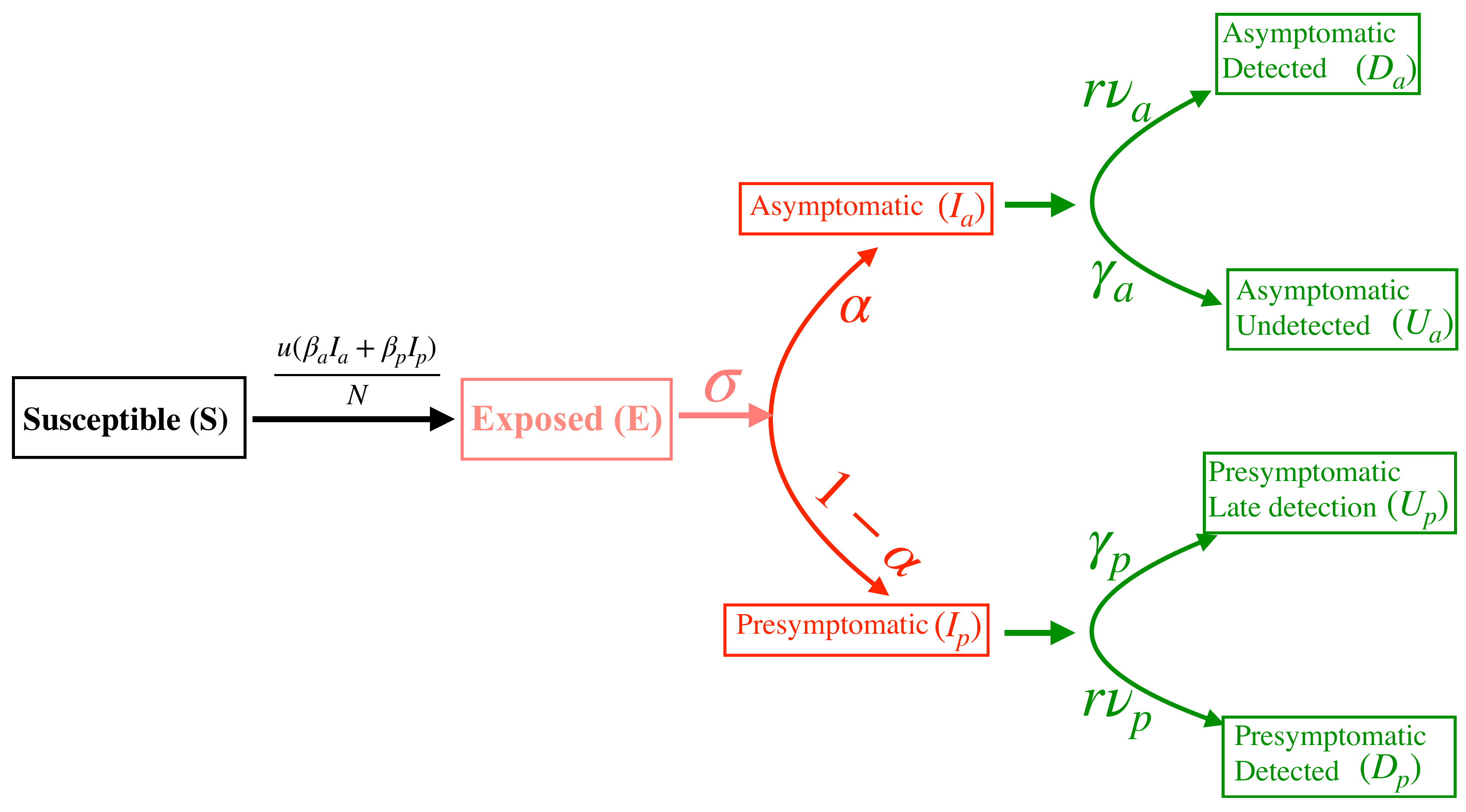}
\caption{A schematic description of the extended SEIR dynamics studied in this work. The parameters $\beta_a,\beta_p,\sigma,\gamma_a,\gamma_p,\alpha$ are intrinsic to the disease, $u$ quantifies the degree of social distancing while  $\nu_a,\nu_p,r$ are related to intervention arising from testing-quarantining.}
\label{schematic}
\end{figure}
The extended SEIR model studied here is schematically described in Fig.~\ref{schematic}. It has eight variables ($S,E,I_a,I_p,U_a,D_a,U_p,D_p$) and ten parameters ($\beta_a,\beta_p,\sigma,\gamma_a,\gamma_p,\alpha,\nu_a,\nu_p,r,u$), of which $\alpha$ represents the fraction of asymptomatic carriers while $u,r,\nu_a,\nu_p$ are related to intervention strategies.

We consider a population of size $N$ that is divided into eight compartments: 

\noindent 1.  $S=$ {Susceptible individuals.} 

\noindent 2.  $E=$ {Exposed but not yet contagious individuals.} 

\noindent 3.  $I_a=$ {Asymptomatic, either develop no symptoms or mild symptoms.} 

\noindent 4.  $I_p=$ {Presymptomatic, those who would eventually develop strong symptoms.}

\noindent 5.  $U_a=$ {Undetected  asymptomatic individuals who have recovered.}

\noindent 6.  $D_a=$ {Asymptomatic individuals who are detected because of directed testing-quarantining, may have mild symptoms, and would have been placed under home isolation (few in India).}

\noindent 7. $U_p=$ {Presymptomatic individuals who are  detected at a late stage after they develop serious symptoms and report to hospitals.  Here we assume that all individuals who develop significant symptoms are eventually detected.}

\noindent 8.  $D_p=$ {Presymptomatic individuals who are detected because of directed testing-quarantining.}

We have the constraint that  $N=S+E+I_a+I_p+U_a+D_a+U_p+D_p$. A standard dynamics for the population classes is given by the following set of equations:
\begin{align}
\f{dS}{dt}&=-\f{u (\beta_a I_a + \beta_p I_p)}{N} S \label{Seqn} \displaybreak[0]\\
\f{dE}{dt}&=\f{u (\beta_a I_a + \beta_p I_p)}{N} S -\sigma E \label{Eeqn} \displaybreak[0]\\
\f{dI_a}{dt}&=\alpha \sigma E - \gamma_a I_a- r \nu_a  I_a \label{Iaeqn}\displaybreak[0] \\
\f{dI_p}{dt}&=(1-\alpha) \sigma E -  \gamma_p I_p -r \nu_p  I_p \label{Ipeqn} \displaybreak[0] \\
\f{dU_a}{dt}&= \gamma_a I_a \label{Uaeqn}\displaybreak[0] \\
\f{dD_a}{dt}&= r\nu_a  I_a \label{Daeqn} \displaybreak[0] \\
\f{dU_p}{dt}&= \gamma_p I_p \label{Upeqn} \displaybreak[0] \\
\f{dD_p}{dt}&= r\nu_p  I_p. \label{Dpeqn} 
\end{align} 
The parameters in the above equations correspond to
\begin{itemize}
\item $\alpha$: fraction of asymptomatic carriers.
\item $\beta_a$: infectivity of asymptomatic carriers.
\item $\beta_p$: infectivity of presymptomatic carriers.
\item $\sigma$: transition rate from exposed to infectious.
\item $\gamma_a$: transition rate of asymptomatic carriers to recovery or hospitalization.
\item $\gamma_p$: transition rate of presymptomatics to recovery or hospitalization.
\item $\nu_a, \nu_p$: detection probabilities of asymptomatic carriers and symptomatic carriers.   
\item $u$: intervention factor due to social distancing (time dependence will be specified later).
\item $r$: intervention factor due to testing-quarantining  (time dependence to be specified later). This is a rate and depends on testing-quarantining rates.
\end{itemize}
With our definitions, the total number of confirmed cases, $C$, and the number of daily recorded new cases $F$ would be 
\begin{align}
C=D_a+D_p+U_p,~F=\f{dC}{dt}=r\nu_a I_a + (\gamma_p+ r \nu_p) I_p~. 
\end{align}
Note that we include $U_p$ because these are people who are not detected through directed tests but eventually get detected (after $\sim 1/\gamma_p$ days) when they get very sick and go to  hospitals. On the other hand the class $D_p$ get detected  because of directed testing, even before they get very sick.

In the next section we will discuss the case where the intervention parameters $u$ and $r$ are kept fixed, and present a number of analytic results. In Sec.~\eqref{sec:interventions} we will discuss the case where  $u$ and $r$ are time-dependent.

\section{Analytic results for model with constant parameters}
\label{sec:analytics}
\subsection{Linear analysis of the dynamical equations}
\label{sec:linear}

Since at early times $S \approx N$ and all the other populations $E,I_a,I_p,D_a,D_p,U_a,U_p \ll N$, one can perform a linearization of the above equations. This tells us about the early time growth of the pandemic, in particular the exponential growth rate. 
Let us define new variables to characterize the linear regime: $x_1=S-N,x_2=E,x_3=I_a,x_4=I_p,x_5=U_a,x_6=D_a,x_7=U_p,x_8=D_p$.  At early times when $x_i << N$, the  dynamics is captured by linear equations
\begin{align}
\f{dX}{dt}&=M X,~~\text{with}~ X=(x_1,x_2,\ldots,x_8), \\
M&=
 \left( \begin{array}{cccccccc} 
0 & 0 & -\tb_a & -\tb_p & 0 &0&0&0 \\
0&-\sigma & \tb_a & \tb_p &0&0&0&0  \\ 
0 & \alpha \sigma & -\tg_a & 0& 0& 0& 0& 0\\ 
0 & (1-\alpha) \sigma & 0& -\tg_p & 0& 0& 0& 0\\ 
0 & 0 &   \gamma_a & 0& 0& 0& 0& 0\\ 
0 & 0&  r\nu_a  & 0&0& 0& 0& 0\\ 
0 & 0 & 0&   \gamma_p & 0& 0& 0 & 0\\ 
0 & 0 & 0& r \nu_p  & 0& 0& 0 & 0
\end{array} \right)~, 
\label{M2linseir}
\end{align}
where $\tb_a=u \beta_a,~\tb_p=u \beta_p, \tg_a=\gamma_a+r \nu_a,\tg_p=\gamma_p+r \nu_p$. For the present we  ignore the time dependence of the SD factor $u$ and the TQ factor $r$.
The matrix has $5$ zero eigenvalues while the  $3$ non-vanishing ones are given by the roots of the following cubic equation for $\lambda$:
\begin{align}
& \lambda^3 
+ (\tg_a +  \tg_p   + \sigma) \lambda^2 
\nonumber\\& 
+ [ \tg_a \tg_p + \tg_a \sigma + \tg_p \sigma -\alpha \tb_a \sigma  - (1-\alpha) \tb_p \sigma  ] \lambda 
\nonumber \\& ~~~ 
+\sigma \left[ \tg_a   \tg_p -(1-\alpha) \tb_p  \tg_a  -\alpha \tb_a     \tg_p  
   \right]=0.
\end{align}
This can be re-written in the form
\begin{align}
 \lambda^3 
+ (\tg_a +  \tg_p   + \sigma) \lambda^2 
+ [ \tg_a \tg_p + \sigma ( \tg_a + \tg_p) (1-Q) ] \lambda  \nn \\
+\sigma  \tg_a   \tg_p (1-R_0)=0, \label{roots}
\end{align}
where $\tb_a=u\beta_a, \tb_p=u \beta_p, \tg_a=\gamma_a +r\nu_a$, $\tg_p=\gamma_p +r\nu_p$,  ${Q} = \alpha {\tb_a}/{(\tg_a+\tg_p)}   + (1-\alpha) {\tb_p}/{(\tg_a+\tg_p)}$, and
\begin{align}
\begin{split}
R_0 &= \alpha \f{\tb_a}{\tg_a} +(1-\alpha) \f{\tb_p}{\tg_p} \\
&=  \alpha \f{u \beta_a}{\gamma_a +r\nu_a} +(1-\alpha) \f{u \beta_p}{\gamma_p +r\nu_p}
\end{split}
 \label{repno}
\end{align}
 is the expected form for the reproductive number for the disease.  One can intuitively see this as follows. The reproductive number is the average number of secondary infection from one infected individual at the initial phase of the outbreak. In our model, an infected individual may either be asymptomatic or presymptomatic with probabilities $\alpha$ or $(1-\alpha)$ respectively. On average, While an asymptomatic individual infects $\bar{\beta_a}/\bar{\gamma_a}$ number of people, a presymptomatic carrier infects $\bar{\beta_p}/\bar{\gamma_p}$ number of people. Consequently the expected number of secondary infection from an arbitrarily chosen infected individual will be $R_0$ with expression given by Eq.~\eqref{repno}.
Noting the fact that $Q <R_0$, it follows that the condition for at least one positive eigenvalue is
\begin{align}
R_0 > 1. 
\end{align}
We  denote the largest eigenvalue by $\mu$ and note that this is uniquely related to the reproductive number $R_0$ by Eq.~\eqref{roots}. At early times the number of cases detected would grow as $\sim e^{\mu t}$. For $R_0 \approx 1$, we expect that the largest eigenvalue is close to zero and,   after neglecting the $\lambda^3$ and $\lambda^2$ terms in Eq.~\eqref{roots}, we can read off the value as
\begin{align}
\mu \approx \f{\sigma (R_0-1)}{1+\sigma (\tg_a^{-1}+\tg_p^{-1}) (1-Q)}.
\end{align}
\\

\noindent
{\bf Initial conditions}:  
We discuss here the fact that all initial conditions (which satisfy the condition $S(0) \approx N$)  quickly move along the direction of the dominant eigenvector and how this provides us a way to choose the correct initial conditions from the knowledge of one variable (e.g confirmed cases)  at an early time.
We denote the right and left eigenvectors corresponding to  an eigenvalue $\lambda_q$ by $\phi_q(i)$ and $\chi_q(i)$ respectively. The largest eigenvalue is denoted by $\mu$ with corresponding right and left eigenvectors  $\phi_m(i)$ and $\chi_m(i)$ respectively.
The vector $X=(x_1,x_2,x_3,x_4,x_5,x_6,x_7,x_8)$ can be written as \cite{Lax}
\begin{equation}
 x_i(t) = \sum_q c_q e^{\lambda_q t} \phi_q(i) \label{eigenExpansion}
\end{equation}  
where the coefficients $c_q$ are determined by putting $t=0$ and then taking the inner product with the left eigenvector say $\chi_q(i)$. Doing so, we get $c_q = \sum_j \chi_q(j) x_j(0)$ which when substituted back in Eq.~\eqref{eigenExpansion} gives
\begin{align}
x_i(t)&= \sum_j \sum_q \phi_q(i) \chi_q(j) e^{\lambda_q t} x_j (0) 
\nn \\
&\approx \sum_j \phi_m(i) \chi_m(j) e^{\mu t} x_j (0), \nn \\
&\approx c_m \phi_m(i) e^{\mu t},~{\rm where}~c_m= \sum_j  \chi_m(j) x_j (0)
\end{align}
where the second last line is true at sufficiently large times when only one eigenvalue $\mu$ dominates. This proves that the direction of the vector $X$ is independent of initial conditions. In particular, using the explicit form of the dominant eigenvector we find the following relation in the growing phase of the pandemic:
\begin{align}
\f{I_a(t)}{I_p(t)}=\f{\phi_m(3)}{\phi_m(4)}=\f{\alpha (\mu+\tilde{\gamma}_p)}{(1-\alpha) (\mu+\tilde{\gamma}_a)}~.
\label{ratio}
\end{align}
Let us consider the initial condition $X=(-\epsilon,0,0,\epsilon,0,0,0,0)$ so that (noting that $\chi_m(1)=0$)
\begin{align}
x_i(t)\approx  \epsilon \phi_m(i) \chi_m(4)  e^{\mu t} = a_i \epsilon e^{\mu t},
\end{align}
where $a_i=\phi_m(i) \chi_m(4)$. 
At a sufficiently large time $t_l$ (but still in the very early phase of the pandemic) we equate the observed confirmed number $C_0$ on some day to $x_6(t_l)+x_7(t_l)+x_8(t_l)$ which therefore gives us the relation
\begin{equation}
\epsilon e^{\mu t_l} = \f{C_0}{a_6 + a_7+ a_8}.
\end{equation}
This then tells us that we should start with the following initial conditions, (now counting time $t=0$ from the day of the observation $C_0$): 
\begin{align}
x_i(0)=\f{\phi_m(i)}{\phi_m(6)+\phi_m(7)+\phi_m(8)}C_0.
\end{align}
\emph{The crucial point is that the leading eigenvector fixes the direction of the growth and then knowledge of linear combination fixes all the other coordinates}.  Thus, independent of initial conditions, the vector describing all the system variables  quickly points along the direction of the eigenvector corresponding to the largest eigenvalue~\cite{frank2020a,frank2020b}. Hence  if we know any one variable (or a linear combination of all the variables) at sufficiently large times in the growing phase, then the full vector is completely specified.  \emph{This leads to an accurate way of specifying initial conditions for the numerics (from insufficient data)} and will help in reducing the number of fitting parameters in modeling studies, thereby  increasing their accuracy in predicting.

\begin{figure}{}
\center
\includegraphics[scale=0.27]{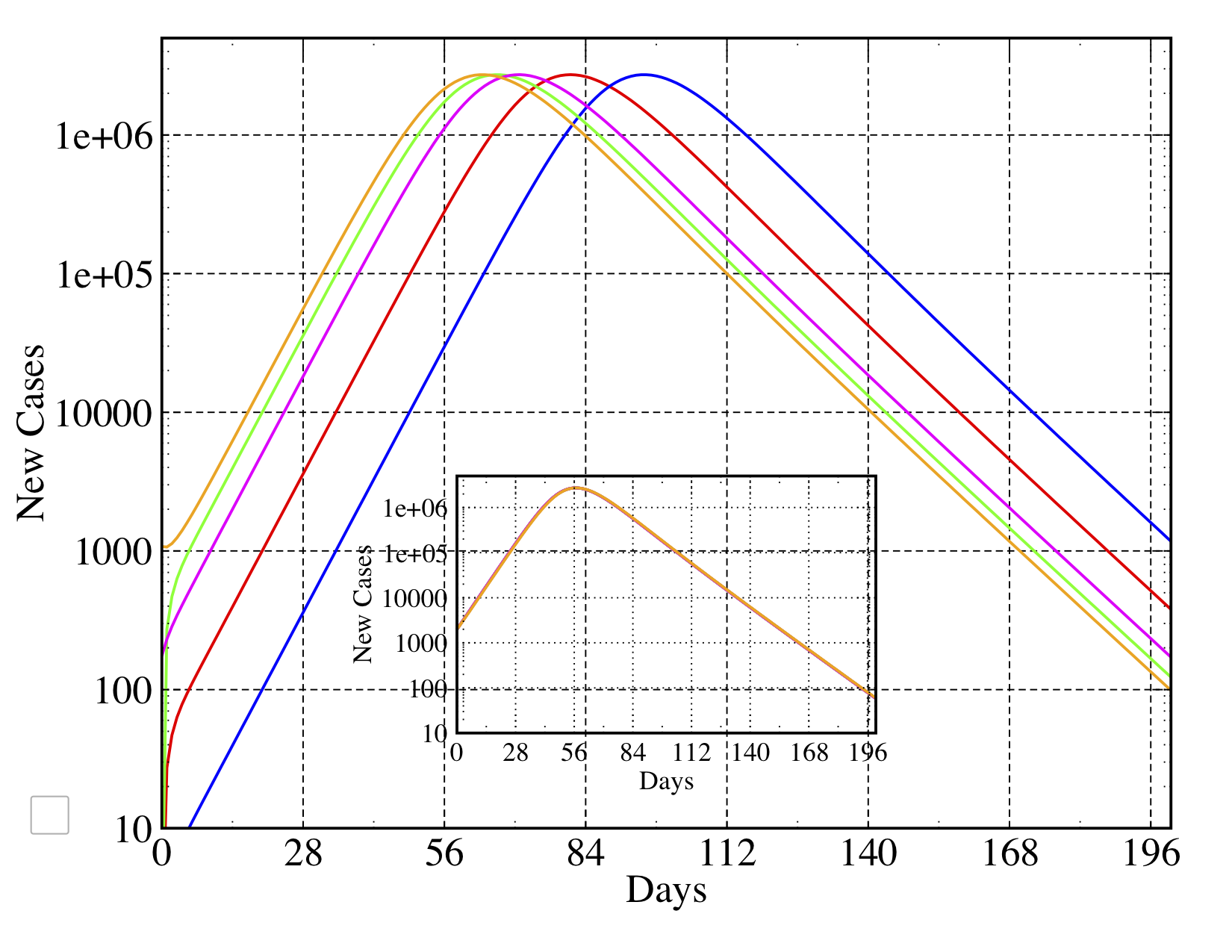}
\caption{{\bf Role of initial conditions}: Plot showing $I(t)$, for a fixed population of size $N=10^7$, with $5$ very different initial conditions : (1)$E(0)=100, I_a(0)=0,I_p(0)=0$, (2)$E(0)=10, I_a(0)=0,I_p(0)=0$, (3)$E(0)=1000, I_a(0)=0,I_p(0)=0$, (4) $E(0)=233, I_a(0)=100,I_p(0)=75$, (5) $E(0)=233, I_a(0)=1000,I_p(0)=75$.  (Inset) A collapse of all the curves obtained by translating all the trajectories so that they start with the same value of $I$.}
\label{init}
\end{figure}
\begin{figure}[]
\center
\includegraphics[width=7.9cm]{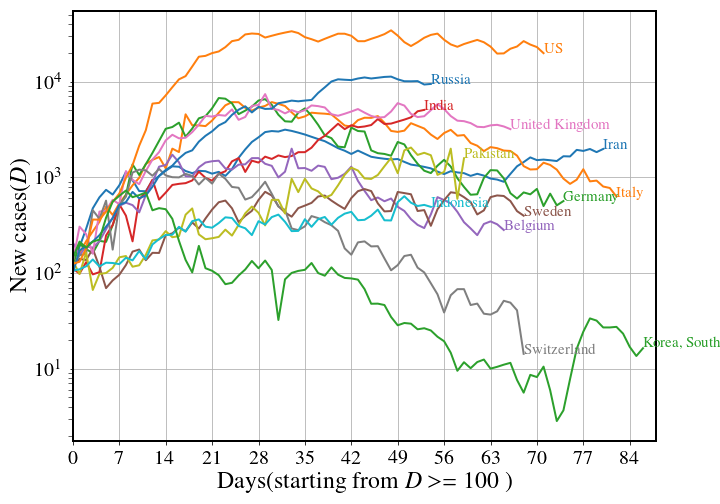}
\includegraphics[width=7.9cm]{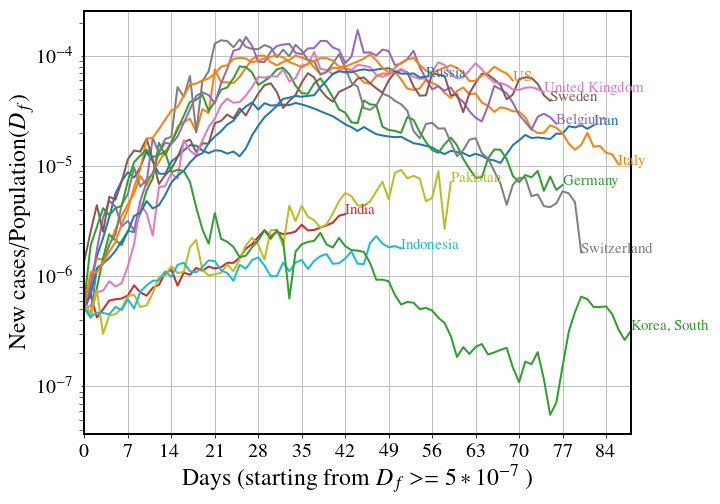}
\caption{ ({\color{blue}left}) Number of new cases per day for different countries. ({\color{blue}right}) Number of new cases normalized by the total population, with the time axis shifted so that every country starts with the same normalized value. Data from \cite{data1}} 
\label{dataI}
\end{figure}

This  fact also implies that different initial conditions (such as different seed infections) will only cause a temporal shift of the observed evolution. 
 This also means that  trajectories  for different initial conditions are identical up to a time translation. 
If one uses identical parameters and intervention strategies, then all countries should follow the same trajectory provided they start with the same value for the normalized fraction of confirmed new cases $F_0/N$.  We illustrate this idea,  for the extended SEIR dynamics, in Fig.~\ref{init} where we show a plot of $I(t)=I_a(t)+I_p(t)$ for $5$ different initial conditions. The inset shows a collapse of all the trajectories after an appropriate time translation of the different trajectories. Can we see a similar collapse of the real data for different countries (after normalizing by the respective populations and with appropriate time translation of the data) ? In Fig.~\ref{dataI} we plot the data with this normalization and initial condition and see a rough collapse for several countries. The differences can be attributed to different parameter values and different control strategies in different countries. We notice in particular that three of the Asian countries (India, Pakistan, Indonesia)  follow a distinctly different trajectory.

\begin{figure*}[t]
\center
\includegraphics[scale=0.24]{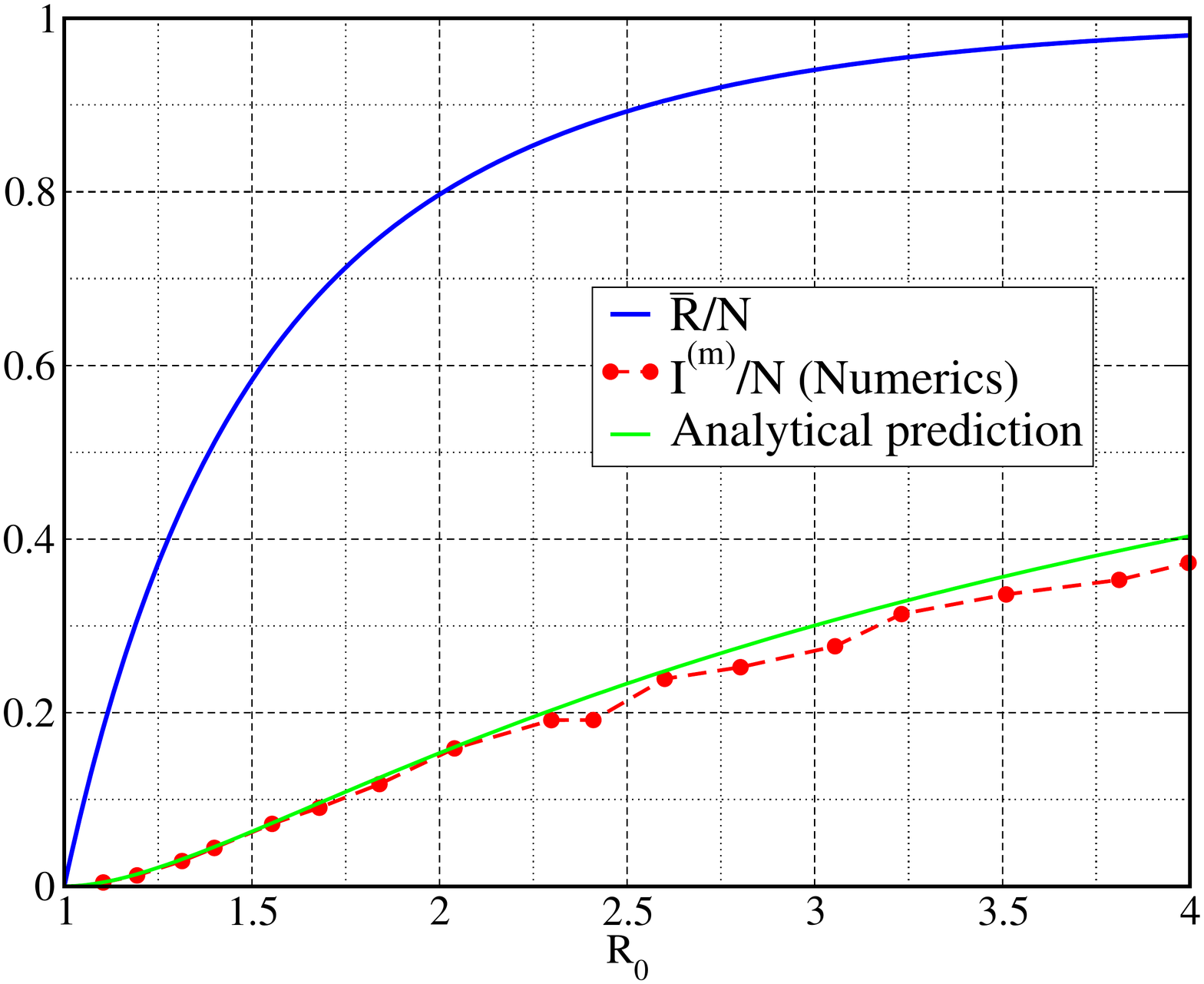}
\includegraphics[scale=0.26]{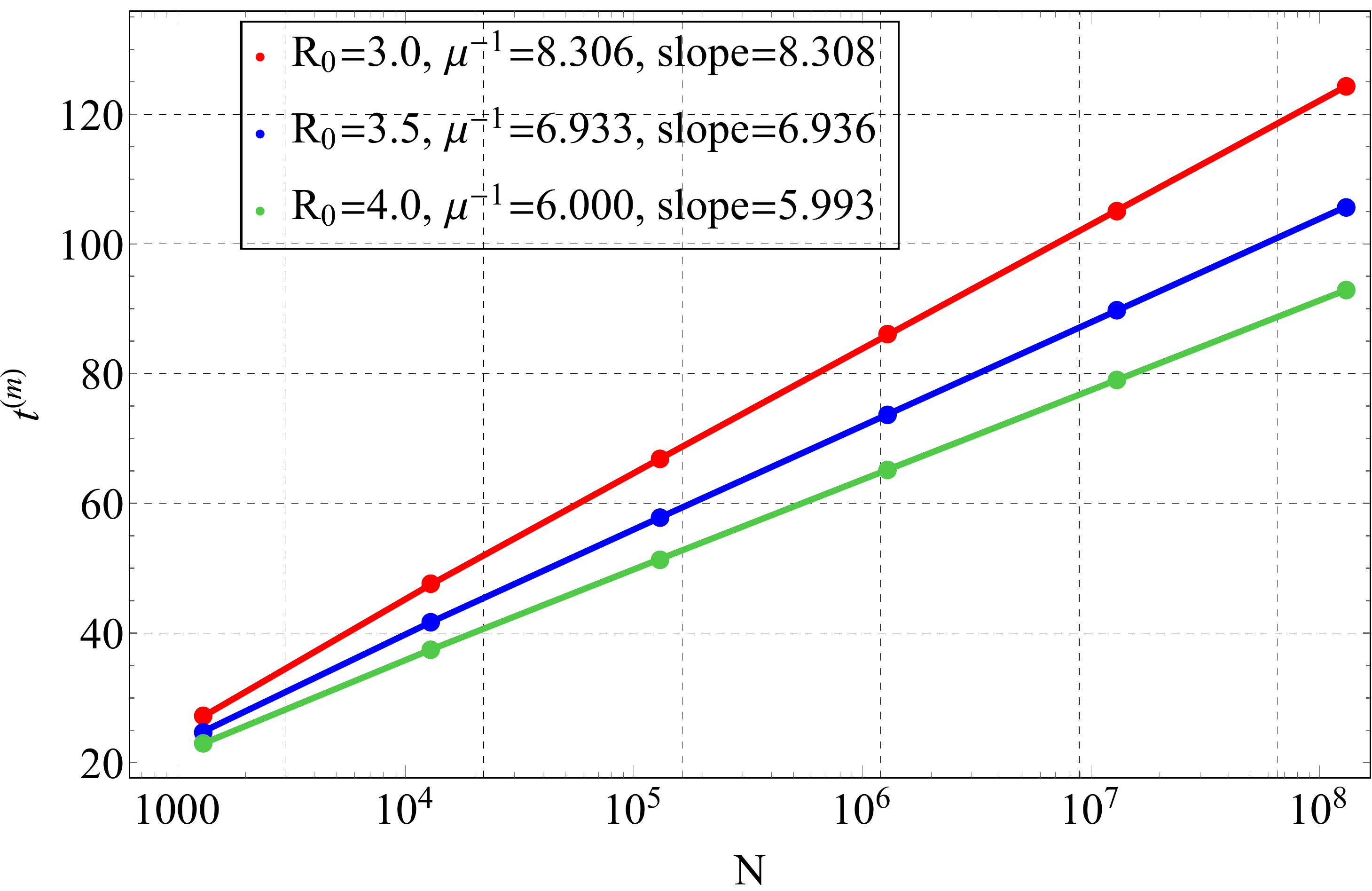}
\caption{ ({\color{blue} left}) Plot of the asymptotic total affected population fraction, $\bar{R}/N$, as a function of the reproductive number $R_0$.  The parameters used are $\beta_a = 0.33, \beta_p = 0.5, \sigma = 0.33, \gamma_a = 0.125, \gamma_p = 0.083, r = 0.0 \;\text{and} \;u = 1.0$.  We also plot the quantity $(I^{(m)}/N) (\sigma+\gamma_e)/\sigma$, obtained numerically from many different parameter sets, and compare it with the theoretical predicted curve $1-(1+\ln R_0)/R_0$ (green line). ({\color{blue} right}) {Verification of the $\ln (N)$ dependence of $t^{(m)}$ in Eq.~\eqref{eq:peakT} for different choices of $R_0$. The slopes of the straight lines compares well with $\mu^{-1}$ as stated in Eq.~\eqref{eq:peakT}.} The parameters used which correspond to Eqs.~\eqref{S_seir}-\eqref{R_seir} are $\sigma = 1/3, \gamma_e = 0.1$ for all values of $R_0$ and $\beta = R_0 \gamma_e$. The dominant eigenvalue $\mu$ are computed using the linearised version of Eqs.~\eqref{S_seir}-\eqref{R_seir} and slopes are obtained from the data.}
\label{Rbar}
\end{figure*}

\subsection{Final  affected population}
\label{subsec:analytic}
Let us define the asymptotic populations (i.e the populations at very long times) in the different compartments as $\bar{U}_a,\bar{D}_a,\bar{U}_p,\bar{D}_p$, and  let $\bar{R}_a=\bar{U}_a+\bar{D}_a,~\bar{R}_p=\bar{U}_p+\bar{D}_p$, $\bar{R}=\bar{R}_a+\bar{R}_p$. The total population that would  eventually be affected by the disease (and either recover or die) is given by $\bar{R}$ and would have developed immunity. A fraction $\bar{U}_a$ (see below) would be undetected and uncounted.

It is possible to compute the final affected population $\bar{R}$ from the dynamical equations in Eqs.~\ref{Seqn} - \ref{Dpeqn}. For the moment let us  assume that $u$ and $r$ do not have any time dependence. We also assume that $U_a(0)=0,~U_p(0)=0,~D_a(0)=0,~D_p(0)=0$ and $S(0) \approx N$. Then solving Eq.~\eqref{Seqn}, we get 
\begin{align}
\bar{S}=N e^{- \int_0^\infty dt [\tilde{\beta}_aI_a(t)+\tilde{\beta}_pI_p(t)]/N}.
\label{S(t)}
\end{align} 
where $\tilde{\beta}_{a}$ and  $\tilde{\beta}_{p}$ are given after Eq.~\eqref{roots}. 
Adding Eqs.~(\ref{Uaeqn}) and (\ref{Daeqn}) and then multiplying both sides by $ \tilde{\beta}_a/\tilde{\gamma}_a$ we get $\frac{ \tilde{\beta}_a}{\tilde{\gamma}_a}\frac{dR_a}{dt} = \tilde{ \beta}_a I_a$ where $R_a=U_a+D_a$. Similarly, we also get 
$\frac{ \tilde{\beta}_p}{\tilde{\gamma}_p}\frac{dR_p}{dt} = \tilde{ \beta}_p I_p$ where $R_p=U_p+D_p$. 
Plugging these two equations into Eq.~\eqref{S(t)} then gives
\begin{align}
\bar{S}=N e^{ -[(\tilde{\beta}_a/\tilde{\gamma}_a) \bar{R}_a/N + 
(\tilde{\beta}_p/\tilde{\gamma}_p)\bar{R}_p/N ] }. \label{Sasym1}
\end{align} 
Next we note that $(d/dt)(I_a+R_a)=\alpha \sigma E$ and $(d/dt)(I_p+R_p)=(1-\alpha) \sigma E$. Hence, for the initial condition $I_a=I_p=R_a=R_p=0$, we find that the ratio $ [I_a(t)+R_a(t)]/[I_p(t)+R_p(t)]=\alpha/(1-\alpha)$ at all times. Since at large times $I_{a,p} \to 0$, this means that the asymptotic values of $R_{a}$ and $R_p$ are given by
\begin{align}
\bar{R_a}= \alpha\bar{R},~~\text{and}~~\bar{R}_p=(1-\alpha) \bar{R}.
\end{align}
Using this in Eq.~\eqref{Sasym1}, noting that $\bar{S}+\bar{R}=N$ and defining $\bar{x}=\bar{R}/N$, we then get the following simple equation that determines the asymptotic total affected population:
\begin{align}
1-\bar{x}=e^{-R_0 \bar{x}},
\label{r0eq}
\end{align}
where $R_0= \alpha \f{\tilde{\beta}_a}{\tilde{\gamma}_a}+(1- \alpha) \f{\tilde{\beta}_p}{\tilde{\gamma}_p}$ is the reproductive number as stated earlier.
 We note that Eq.~\eqref{r0eq} has a non-zero solution only when $R_0 >1$. In Fig.~\eqref{Rbar} we show the dependence of $\bar{x}$ on $R_0$, obtained from a numerical solution of Eq.~\eqref{r0eq}.  For the simple SIR model the result of Eq.~\eqref{r0eq} is well known \cite{book}, here we show that this is valid for an extended model as well generally. This computation of the asymptotic  population can be straightforwardly  extended to a more general model where one can have arbitrary number of compartments for the infected and recovered populations.

The asymptotic population of all four compartments within R are thus given by
\begin{align}
\bar{R}_a&=\alpha \bar{R}, \;\bar{R}_p=(1-\alpha) \bar{R} \nn \\ 
\bar{U}_a&=\alpha\f{\gamma_a}{\gamma_a+r \nu_a}\bar{R}, \;\bar{D}_a=(1-\alpha)\f{r \nu_a}{\gamma_a+r \nu_a}\bar{R}, \nn \\
\bar{U}_p&=\alpha\f{\gamma_p}{\gamma_p+r \nu_p}\bar{R}, \;\bar{D}_p=(1-\alpha)\f{r \nu_p}{\gamma_p+r \nu_p}\bar{R}. 
\end{align}

\subsection{Peak infections and the time to reach the peak}
\label{peak-I}
The peak infection numbers and the time at which the peak occurs are two important quantities that one would like to know during a pandemic. Here we provide heuristic analytic formulas for these, which are derived after making some simple physical assumptions.   

We start with the evolution equations for the infected populations, given by:
\begin{align}
\frac{d I_a}{dt} &= \alpha \sigma E - \tilde{\gamma}_a I_a, \nn \\
\frac{d I_p}{dt} &= (1-\alpha) \sigma E - \tilde{\gamma}_p I_p. \label{dI}
\end{align} 
Let us assume that  in comparison to the time scale of the pandemic, $E, \, I_a$ and $I_p$ all peak at  roughly the same time say $t_m$, and let us indicate by $E^{(m)}, I_a^{(m)}, I_p^{(m)}$ the respective peak values. We then obtain 
\begin{align}
\tilde{\gamma}_a I_a^{(m)} &= \alpha \sigma E^{(m)} \nn \\
\tilde{\gamma}_p I_p^{(m)} &= (1-\alpha) \sigma E^{(m)}. \label{I_ap-E}
\end{align} 
Defining $I^{(m)} = I_a^{(m)} + I_p^{(m)}$, Eq.~\eqref{I_ap-E} gives 
\begin{align}
E^{(m)} &= \frac{\gamma_e}{\sigma} I^{(m)}, \label{E-I_ratio} \\
{\rm where}~
\gamma_e &= [\alpha \tilde{\gamma}_a^{-1} + (1-\alpha) \tilde{\gamma}_p^{-1}]^{-1} \label{gammaeff}
\end{align} 
is an effective recovery rate. Substituting this in Eq.~\eqref{I_ap-E} yields
\begin{align}
I_a^{(m)} &= \alpha\frac{\gamma_e}{\tilde{\gamma}_a} I^{(m)} \nn \\ 
I_p^{(m)} &= (1-\alpha)\frac{\gamma_e}{\tilde{\gamma}_p} I^{(m)}. \label{I_ap^max}
\end{align}
Let us further assume that Eq.~\eqref{I_ap^max} is valid in general (for all time) and not just at the peak. Then, after defining  $I=I_a+I_p$,  we have
\begin{align}
I_a = \alpha\frac{\gamma_e}{\tilde{\gamma}_a} I,~~I_p = (1-\alpha)\frac{\gamma_e}{\tilde{\gamma}_p} I. \label{I_ap^gen}
\end{align}
Substituting Eq.~\eqref{I_ap^gen} in the model equations \eqref{Seqn}-\eqref{Dpeqn}, and defining $R=U_a+D_a+U_p+D_p$, we obtain the standard SEIR equations,
\begin{align}
 \f{dS}{dt} &= -\f{\beta_e I}{N}S, \label{S_seir}\\
 \f{dE}{dt} &= \f{\beta_e I}{N}S - \sigma E, \label{E_seir}\\
 \f{dI}{dt} &= \sigma E - \gamma_e I, \label{I_seir}\\
 \f{dR}{dt} &= \gamma_e I, \label{R_seir}
\end{align}
Here, $\gamma_e$ is given by Eq.~\eqref{gammaeff} and $\beta_e = R_0 \gamma_e$ where the reproductive number $R_0$ is given by Eq.~\eqref{repno}. Assuming the initial conditions: $S(0) \approx N, R(0)=0$, we can  solve Eqs.~(\ref{S_seir},\ref{R_seir}) to get:
\begin{align}
S &= N e^{-R_0 R/N}. \label{S_soln}
\end{align} 
At the peak, our assumption of $dE/dt=0$ gives from Eq.~\eqref{E_seir},
$S^{(m)}/N=\sigma E^{(m)}/(\beta_e I^{(m)}) = \gamma_e/\beta_e=R_0^{-1}$ [using Eq.~\eqref{E-I_ratio}].  The peak value  $R^{(m)}$ is then given from Eq.~\eqref{S_soln} by $R^{(m)}/N=R_0^{-1}\ln (R_0)$.  
Finally, using Eq.~\eqref{E-I_ratio} and the fact that $S/N+E/N+I/N+R/N=1$ gives
\begin{align}
I^{(m)} = \frac{\sigma}{\gamma_e+\sigma}\left(1-\frac{1+\log R_0}{R_0}\right)N. \label{eq:peakI} 
\end{align}
 For somewhat simpler models it is possible to obtain an analytical estimate of the peak size~\cite{piovella2020} and for the time, $t^{(m)}$, required to reach the infection peak~\cite{lou2016}. In our case we estimate $t^{(m)}$  by noting that  the linearized dynamics is approximately valid (see previous section) up to the time $I(t)$ reaches its peak $I^{(m)}$.
Hence we write $I^{(m)}=I_0 \,e^{\mu (t^{(m)}-t_0)}$, where $I_0=I(t=t_0)$ is the infection number at some early  time (but already in the exponentially growing regime).  Hence we get 
\begin{align}
t^{(m)} - t_0 = \frac{\ln[I^{(m)}/I_0]}{\mu} \sim \f{\ln (N/c)}{\mu}, \label{eq:peakT}
\end{align}  
with $I^{(m)}$ given by Eq.~\eqref{eq:peakI}  and $c$ is a constant that depends on initial conditions and model parameters. 

\begin{figure}
\center
\includegraphics[scale=0.5]{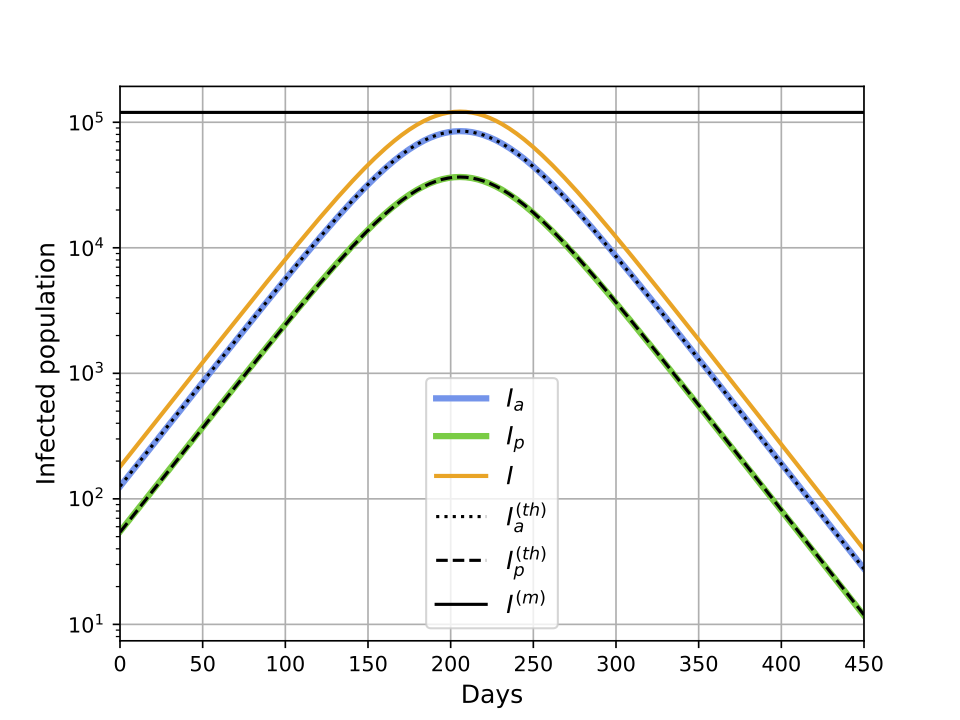}
\caption{ Comparison of values of $I_a,I_p, I$ obtained from direct numerics with those  from the approximate assumed forms, $I_a^{\rm th},I_p^{\rm th}$ from  Eq.~\eqref{I_ap^gen} ($I$ taken from the numerics). We also plot the predicted  peak value of $I^{(m)}$. The parameters in the numerics were taken as  $N=10^7,\alpha=0.67, \beta_a=0.333, \beta_p=0.5, \gamma_a=1/8, \gamma_p=1/12, \sigma=1/3, \nu_a=1/3, \nu_p=1/2, u=1, r=0.5$, which gives $R_0=1.26, \mu=0.039$. We see in this case the assumption is valid to a high degree of accuracy.}
\label{ratiocheck}
\end{figure}

In Fig.~\eqref{Rbar}, we  provide a numerical verification of the results in   Eq.~\eqref{eq:peakI}  and Eq.~\eqref{eq:peakT} by solving the extended SEIR equations numerically. One of the main assumptions required for the proof is the validity of Eqs.~\eqref{I_ap^gen}. In Fig.~\eqref{ratiocheck} we check this assumption in a numerical example with a particular parameter set. 

\section{Interventions: Social distancing and Testing-Quarantining}
\label{sec:interventions}
We next consider the effect of different intervetion strategies which are incorporated into the extended SEIR dynamical equations, \eqref{Seqn}-\eqref{Dpeqn}, through the parameters $u$ and $r$ which we will now make time-dependent. 
We discuss here the choices of the intervention functions $u$ and $r$. Note that $u$ is a dimensionless number quantifying the level of social contacts, while $r$ is a rate which, as we will see, is closely related to the testing rate.  \\

\noindent
{\bf Social distancing (SD)}: We multiply the constant factors $\beta_{a,p}$ by the time dependent function, $u(t)$, the ``lockdown'' function that incorporates the effect of social distancing, i.e reducing contacts between people. A reasonable form is one where $u(t)$ has the constant value $(=1)$  before the beginning of any interventions,  and then from time $t_{on}$ it changes to a value $ 0< u_l <1$, over a characteristic time scale  $\sim t_w$. Thus we take a form
\begin{align}
u(t)&=1~~~t<t_{\rm on}, \nn \\ 
&=u_l+(1-u_l)e^{-(t-t_{\rm on})/t_w},~~~t>t_{\rm on}. 
\end{align}
 The number $u_l$ indicates the lowering of social contacts. \\

\noindent
{\bf Testing-quarantining (TQ)}: We expect that  testing and quarantining will take out individuals from the infectious population and this is captured by the terms $ r \nu_a I_a$ and $ r \nu_p I_p$ in the dynamical equations.  A reasonable choice  for the TQ function is perhaps  to take 
\begin{align}
r(t)&=0~~~t<t'_{\rm on}, \nn \\ 
&=r_l-r_le^{-(t-t'_{\rm on})/t'_w},~~~t>t'_{\rm on}.
\end{align}
where one needs a final rate  $r_l>0$. In general the time at which the TQ begins to be implemented $t'_{\rm on}$ and the time required for it to be effective $t'_{w}$ could be different from those used for SD. 

 A  useful quantity to characterize the system with interventions is the time-dependent effective reproductive number given by 
\begin{align}
R_0^{\rm eff}(t)= \alpha \f{u(t)\beta_a}{\gamma_a +r(t) \nu_a} +(1-\alpha) \f{u(t) \beta_p}{\gamma_p+r(t) \nu_p}.  \label{trepno}
\end{align}
At long times this goes to the targeted reproduction number 
\begin{align}
R_0^{\rm target}=R_0^{\rm eff}(t \to \infty)= \alpha \f{u_l\beta_a}{\gamma_a +r_l \nu_a} +(1-\alpha) \f{u_l \beta_p}{\gamma_p+r_l \nu_p}. \label{tarrep}
\end{align}
The time scale for the intervention target to be achieved is given by $t_w$ and $t'_w$. \\ \\

\begin{figure}[t]
\center
\includegraphics[scale=0.3]{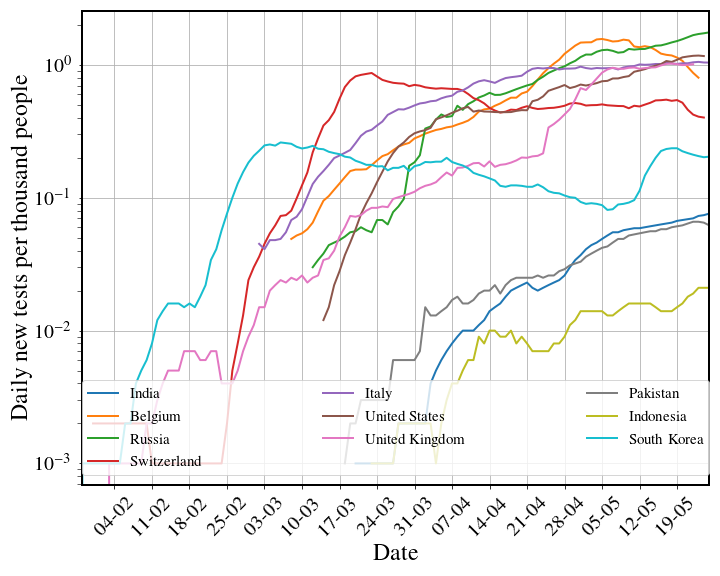}
\caption{Data of number of tests per day per thousand in several countries on a  log-scale.   Data from \cite{data2}.}
\label{testdata}
\end{figure}

\noindent
{\bf Relation of the TQ function $r(t)$ to the number of tests done per day}:
Let us suppose that the number of tests per person per day is given by $T_r$. We show in Fig.~(\ref{testdata}) the data for the number of tests per $1000$ people per day across a set of countries and see that this is around $0.05$ for India which means that $T_r=0.00005$. If tests are done completely randomly, then the number of detected people (assuming that the tests are perfect) would be $T_r \times I$ and so it is clear that we can identify $r(t)=T_r(t)$. It is then clear that this would have no effect on the pandemic control. To have any effect we would need $r \gtrsim \gamma_p \approx 0.1$ which means around $100$ tests per $1000$ people per day which is clearly not practical.

\begin{figure*}[t]
\center
\includegraphics[scale=0.33]{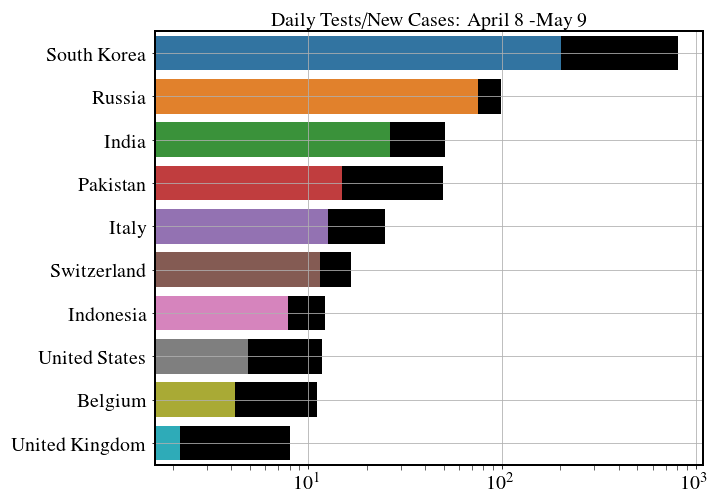}
\includegraphics[scale=0.35]{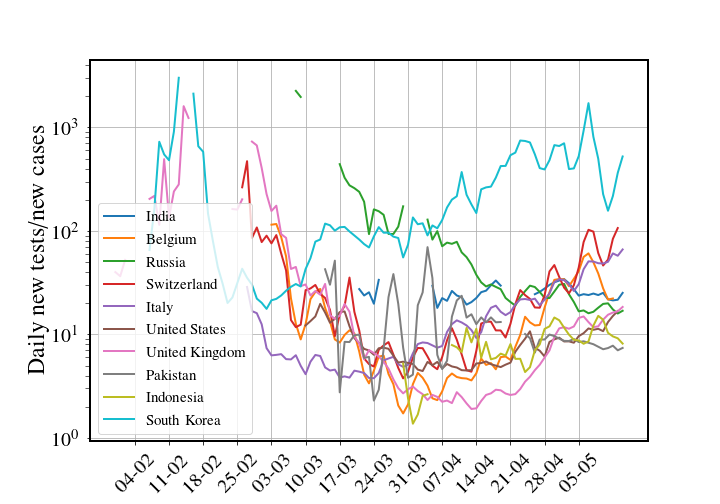}
\caption{Data from different countries on the number of tests per detected case $T(t)/F(t)$ for $10$ countries  ({\color{blue}left}) on the dates April 8 (coloured bar), May 9 (black bar) and  ({\color{blue}right}) the change over time of this ratio. Data from \cite{data2}}
\label{TbyD}
\end{figure*}

However, a better strategy is to do focused tests on the contacts of all those who have been detected on a given day. We now give an estimate of the rate $r$ if we followed this strategy. For simplicity of presentation of our argument we here assume $\nu_a=\nu_p=1$ and $\gamma_a=\gamma_p$.  
From our extended SEIR model the number of detected cases per day is given by $ F(t)=r \nu_a I_a+ (r \nu_p+\gamma_p) I_p = r I+\gamma_p I_p $. In the growing phase we have, from Eq.~\eqref{ratio}, that $I_a=\alpha I$ and $I_p=(1-\alpha) I$. Hence we get $ F(t) = \hat{\gamma} I$ with $\hat{\gamma}= r +(1-\alpha)\gamma_p$. 
The total number of contacts of the $I=F(t)/\hat{\gamma}$ individuals would be $A F(t)/\hat{\gamma}$, where $A$ is the mean number of contacts of a single infected person. If we perform $T$ tests per day \emph{on this pool}, then  the rate of detections will be given by 
\begin{align}
r=\f{T \hat{\gamma}}{A F(t)} \label{Trate1}
\end{align}
Denoting $c=T/(A F)$ and noting that  $\hat{\gamma}= r +(1-\alpha)\gamma_p$, we self-consistently solve the above equation to find
\begin{equation}
r=\f{c(1-\alpha)\gamma_p}{1-c}.  \label{Trate2}
\end{equation}
Now it is clear that unless $r$ and $\gamma_a=\gamma_p$ are of the same  order, TQ will not have much effect on the dynamics and the change in $R_0$ will be small. Setting $r \gtrsim \gamma_p$ then gives us the condition
\begin{align}
{T} \gtrsim \frac{ A F(t)}{2-\alpha}.  \label{Trate3}
\end{align}
Note that in our model we identify $r(t)$ as our control rate function that changes from the value  $0$ to a value  $r_l \approx \gamma_p$ over the time scales of a week or so.    This  means that we would need to change the testing rate in a controlled way such that the condition $T(t) \sim  A F(t)$ is maintained.  Thus \emph{the number of tests/per day  has to be proportional to number of new detections/per day and in fact the ratio $T/F$  has to be larger than the average number of contacts, $A$, that  each infectious person makes.}  The number $A$ is expected to depend on the population density and also how well SD is being implemented. 
 The table in Fig.~\ref{TbyD} shows data for the  ratio $T(t)/F(t)$ for a set of countries and also how this ratio has evolved over time.
While the value of $T(t)/F(t) \approx 25$ (around May 15) for India appears to be large, it may not be sufficient given that the population densities are much larger than in many other countries and implementation of SD may be less effective.  If we assume $20$ contacts a day and the number of days before isolation of the individual to be $5$ we get  the rough estimate of $A \approx 100$ and then the ratio $T/F$ thus has to be at least $\approx 100$. This is the minimum value of testing-to-detected ratio that has to be targeted at  localities with high infection rates.    The details of the  arguments  presented here are largely \emph{independent of the specifics of the particular SEIR model that we study}.

\subsection{Comparision of intervention strategies}
\label{sec:results}

  The model details are given  in Sec.~\eqref{sec:model}. We recall that at any given time the total  infectious population size is $I=I_a+I_p$, the cumulative affected population (recovered, in hospital or dead) is $R=U_a+D_a+U_p+D_p$, the reported total confirmed cases is $C=D_a+D_p+U_p$, and the reported new daily cases is $F={dC}/{dt}=r\nu_a I_a + (\gamma_p+ r \nu_p) I_p$.

\begin{figure*}[t]
\center
\includegraphics[scale=0.325]{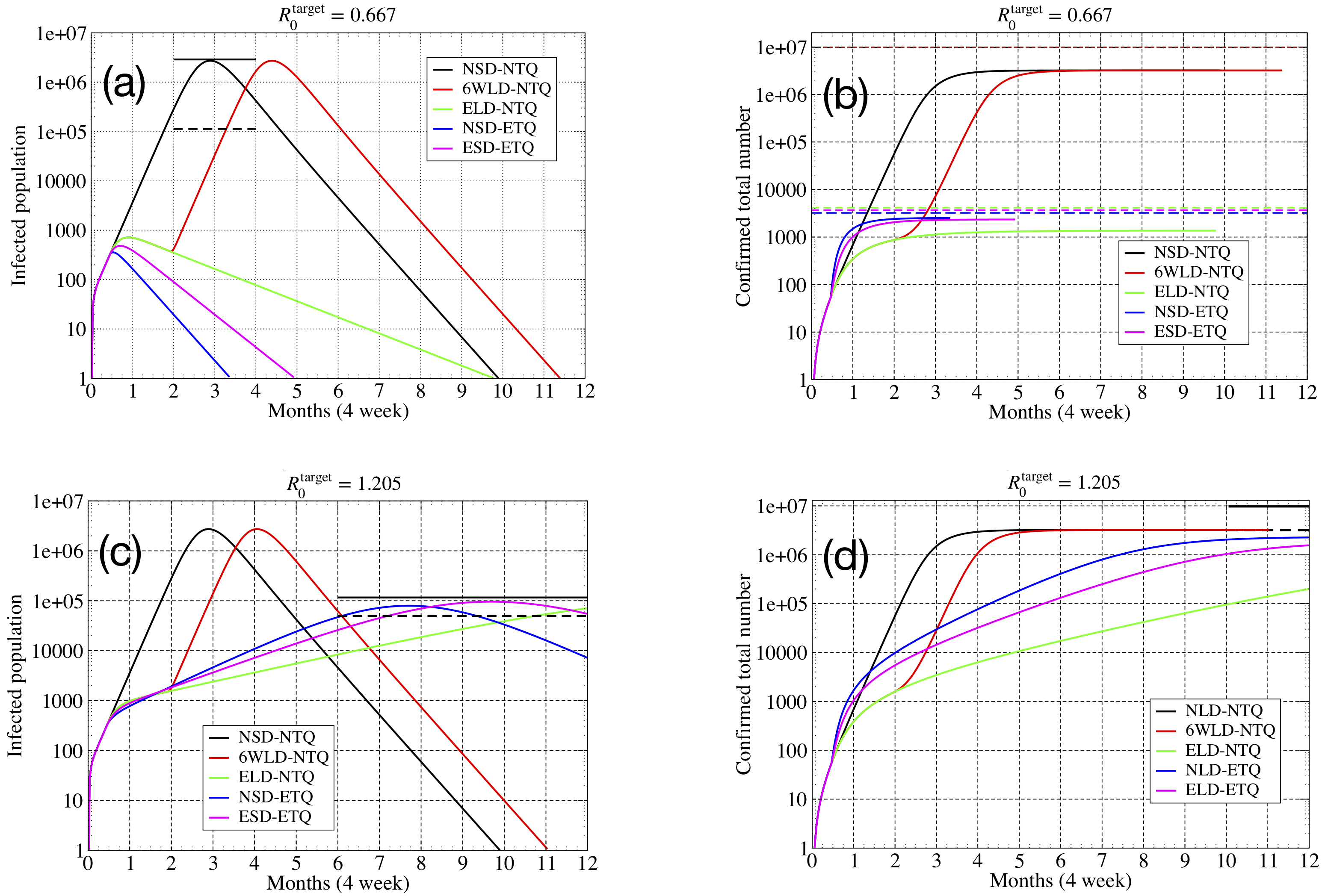}
\caption{{\bf Parameter set I  [$R_0^{\rm target}=0.667$]}: {\color{blue} (a)} Total number of infected cases $I=I_a+I_p$ for different intervention strategies. The solid and dashed black lines indicate the peak infected cases $I^{(m)}$ as given by Eq.~\eqref{eq:peakI} and the 
corresponding value of $I_p^{(m)}$. 
{\color{blue} (b)} Total number of confirmed cases $C=U_p+D_a+D_p$. The dashed lines indicate the total affected population $R=C+U_a$ at the end of one year, for the different strategies. In the absence of interventions  this is close to
$96\%$ and is given by Eq.~\eqref{r0eq}. 
The total population was taken as $N=10^7$. {\bf Parameter set II [$R_0^{\rm target}=1.205$]}:  {\color{blue} (c)} Total number of infected cases $I=I_a+I_p$ for different intervention strategies. {\color{blue} (d)} Total number of confirmed cases $C=U_p+D_a+D_p$. The dashed lines indicate the total affected population $R=C+U_a$ at the end of one year, for the different strategies. Total population was taken as $N=10^7$.}
\label{Pset1}
\end{figure*}

 A  useful quantity to characterize the system with interventions is the 
 targeted reproduction number [see Eqs.~(\ref{trepno},\ref{tarrep})]
\begin{align}
R_0^{\rm target}=  \f{\alpha u_l\beta_a}{\gamma_a +r_l \nu_a} + \f{(1-\alpha) u_l \beta_p}{\gamma_p+r_l \nu_p}.
\end{align}
We classify intervention strategies by the targeted $R_0^{\rm target}$ value. 
A {\bf strong intervention} is one where $ R_0^{\rm target} < 1 $ and will achieve suppression of the disease while a {\bf weak intervention}  is one  with  $ R_0^{\rm target} \gtrsim 1$ and will only mitigate the effects of the disease. 

Other than $R_0$, an important quantity to characterize the disease growth is the largest eigenvalue $\mu$ of the linearized dynamics (see Sec.~\ref{sec:linear}). In the early phase of the pandemic, all populations other than $S$ grow exponentially with time as $\sim e^{\mu t}$. As we will see, for the case of strong intervention, $\mu$ becomes negative and gives the exponential decay rate of the disease.

In our numerical  study we choose, for the purpose of  illustration,  the following parameter set:  $\alpha=0.67$ and the rates $\beta_a=0.333, \beta_p=0.5,\sigma=1/3,\gamma_a=1/8,\gamma_p=1/12$ all in units of day$^{-1}$. For the specified choice of parameter values (free case with $u=1.0, r=0.0$) we get $\mu=0.158$ which is close to the  value observed for the early time data for confirmed cases in India. The corresponding free value of  $R_0$  is $3.7665$. 
Note that $\mu$ is not uniquely fixed by $R_0$ (and vice versa) and different choices of parameters can give the same observed $\mu$ but different values of $R_0$

Choosing these typical parameter values for COVID-19, we now compare the efficacy of strong and weak interventions implemented in four different ways: 
(1) $6$WLD-NTQ: Six weeks lockdown (strong value of SD parameter) and no testing-quarantining, (2) ELD-NTQ: Extended lockdown and no testing-quarantining,(3)NSD-ETQ: No social distancing and extended testing-quarantining, (4) ESD-ETQ: Extended social distancing and  extended testing-quarantining. The case with no social distancing and no testing-quarantining is indicated as NSD-NTQ.

We work with a population $N=10^7$ and initial conditions $E(0)=100$, $I_a(0)=I_p(0)=U_a(0)=D_a(0)=U_p(0)=D_p(0)=0$ and $S(0)=N-E-I_a-I_p-U_a-D_a-U_p-D_p$. In all cases, we will assume that intervention strategies are switched on when the confirmed number of cases reaches $50$ and after that the full intervention values are attained over a time scale of $5$ days.

\subsubsection{Strong intervention ($R_0^{\rm target} <1$)}
 In this case, the exponential growth stops around the time  when  $R_0^{\rm eff}(t)$ crosses the value $1$. After this time, the infection numbers will start decaying exponentially. Since the infection numbers are still small compared to the total population, one can work with the linearized theory and the magnitude of the largest eigenvalue $\mu$ (now negative)  determines the exponential decay rate. For illustrating this case, we take:
\\

\noindent 
{\bf Parameter set I [$R_0^{\rm target}=0.667$]} --- We choose three SD and TQ strengths as
(i) SD: $u_l=0.177, r_l=0$, (ii)  TQ: $u_l=1, r_l=1.2$ and (iii) SD-TQ: $u_l=0.461, r_l=0.4$. This choice corresponds to changing the free value of  $R_0=3.766$ to a  target value $R_0^{\rm target}=0.667$, for all the three different strategies. The largest eigenvalue $\mu$ changes from the free value $\mu=0.158$ to the values (i) $\mu=-0.027$, (ii) $\mu = -0.077$ (iii) $\mu =-0.0546$ respectively.
The results of the numerical solution of the  extended SEIR equations are presented Figs.~(\ref{Pset1}a) and (\ref{Pset1}b). 
\\ \\
\noindent
{\bf Main observations}: 
\begin{enumerate}
\item A six week (or eight week) lockdown is insufficient to end the pandemic and will lead to a second wave. If the interventions are carried on indefinitely,  the pandemic is  suppressed and only affects a very small fraction of the population (less than $0.1\%$). We can understand all features of the dynamics from the linear theory.  In Figs.~\ref{Pset1}(a,b), intervention is switched on after $\approx 2$ weeks and the peak in infections appears roughly after a period of $5$ days. Thereafter however, the decay in the number of infections occurs slowly, the decay rate being given by the largest eigenvalue $\mu$ (now negative and smaller in magnitude than $\mu$ in the growth phase).

\item \emph{We find that for the same target $ R_0^{\rm target} < 1 $, different intervention schemes  (ELD-NTQ, NSD-ETQ, or ESD-ETQ)  can give very different values of the decay rate $\mu$ and, in general we find that TQ is more effective than SD}.  We see that ELD-NTQ ends the pandemic in about $10$ months while NSD-ETQ would take around $3.5$ months. This can be  understood from the fact that the corresponding $\mu$ values (post-intervention) are given by $\mu=-0.027$ and $\mu=-0.077$ respectively, i.e, they differ by a factor of about $3$.  With a mixed strategy where one allows almost three times more social contacts ($u_l=0.431$) than for LD case and that requires three times less testing ($r_l=0.4$) than for TQ case, we see that the disease is controlled in about $5$ months. Hence this appears to be the most practical and effective strategy.

\item The expected time for the pandemic to die would be roughly given by 
\begin{align}
t_{{\rm end}} \sim \f{\ln ({\rm Peak~infection~number})}{|\mu^{\text{post}-\text{intervention}}|},
\end{align}
and so it is important that intervention schemes are implemented early and as strongly as possible.
 \end{enumerate} 

\subsubsection{Weak intervention ($R_0^{\rm target} \gtrsim 1$)}
  In this case,  a finite fraction of the population is eventually affected, but the intervention succeeds in reducing this from its original free value and in delaying considerably the date at which the infections peak. We take the following parameter set for this study:  
\\

\noindent{\bf Parameter set II [$R_0^{\rm target}=1.205$]} --- we choose three SD and TQ strengths as
(i) SD: $u_l=0.32, r_l=0$, (ii)  TQ: $u_l=1, r_l=0.536$ and (iii) SD-TQ: $u_l=0.634, r_l=0.24$. This choice corresponds  to changing the free value of  $R_0=3.766$ to a fixed target value $R_0^{\rm target}=1.205$ for all the three different strategies. The largest eigenvalue $\mu$  remains positive and changes from the free value $\mu=0.158$ to the values (i) $\mu=0.0152$, (ii) $\mu = 0.032$ (iii) $\mu = 0.0248$ respectively. The results of the numerical solution of the  extended SEIR equations are presented Figs.~(\ref{Pset1}c) and (\ref{Pset1}d).
\\ 

{\bf Main observations}:  
\begin{enumerate}
\item 
 We find that in this case the peak infections, peak infection time and the final affected population can be obtained from  the analytic expressions  given by Eqs.~(\ref{r0eq},\ref{eq:peakI},\ref{eq:peakT}) in terms of the basic disease parameters, using their values  after interventions are introduced. We assume that the intervention parameters change from the values $u=0,r=0$ to their full strength $u=u_l,r=r_l$ over a short time scale and thereafter remain constant.

\item We find that the peak infection numbers are smallest for the case with ELD-NTQ and they occur at a later stage. These results can also be understood  mathematically from the  expressions  in Eq.~\eqref{eq:peakI} and Eq.~\eqref{eq:peakT} using the  post-intervention values of $\gamma$ and $\mu$ (from the linear theory).  
\item{We note that while weak interventions can slow down and reduce the impact of the pandemic, they do not lead to development of herd immunity of the population} (assuming that all the recovered people develop immunity). It is well known that herd immunity is attained when a fraction $1-1/R_0$ of the population has developed immunity. Thus herd immunity in the above example would require that $1-R_0^{-1} \approx 0.74$, i.e $74\%$  of the population be affected, while  Eq.~\eqref{r0eq} with $R_0^{\rm target}=1.205$ predicts that only about $31\%$ of the population is affected. 
\end{enumerate}

\begin{figure*}[]
\center
\includegraphics[height=4.cm,width=5.5cm]{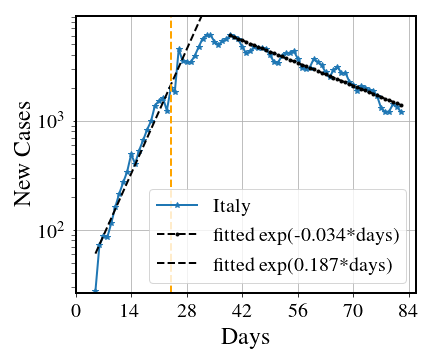}
\includegraphics[height=4.cm,width=5.5cm]{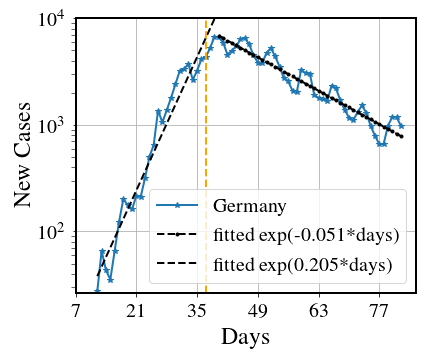}
\includegraphics[height=4.cm,width=5.5cm]{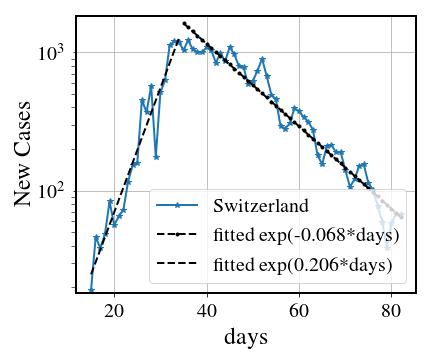}
\includegraphics[height=4.cm,width=5.5cm]{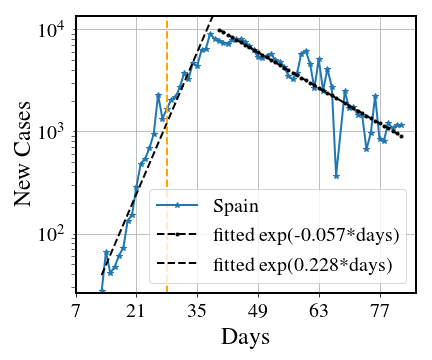}
\includegraphics[height=4.cm,width=5.5cm]{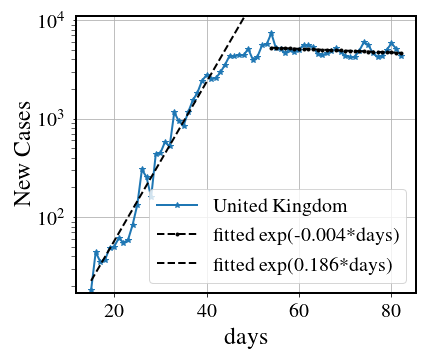}
\includegraphics[height=4.cm,width=5.5cm]{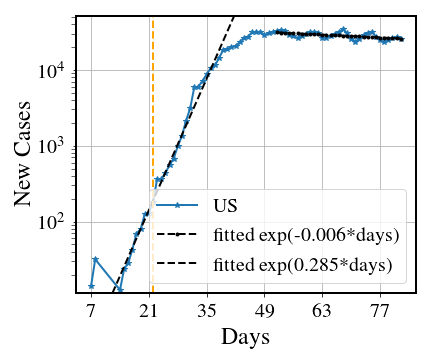}
\includegraphics[height=4.cm,width=5.5cm]{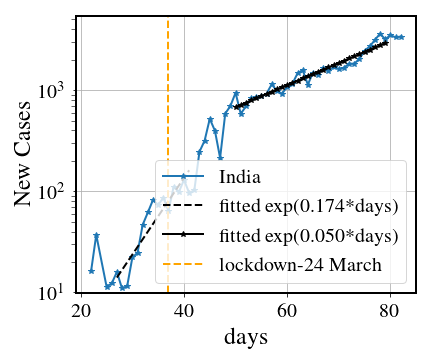}
\includegraphics[height=4.cm,width=5.5cm]{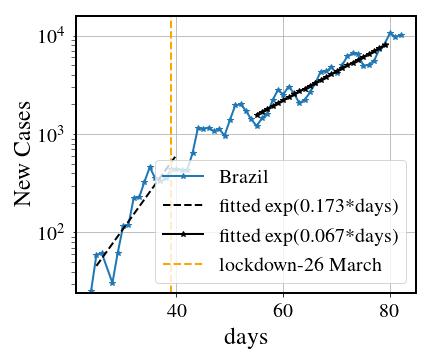}
\includegraphics[height=4.cm,width=5.5cm]{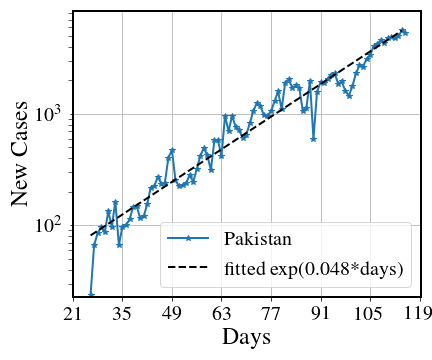}
\caption{ Number of new cases per day for nine different countries. The dashed orange vertical lines in some of the figures denote the day of the implementation of lockdown.  We note that the first six data sets exhibits the same broad features that we see for the model predictions in Fig.~\ref{Pset1}(a,b). In particular we see  the fast exponential growth  and slow exponential decrease in new cases (following strong interventions). The two countries UK and US show a very slow decay rate, indicating that disease suppression has barely been achieved. The data for India, Brazil and Pakistan show the behavior corresponding to model predictions in Fig.~\ref{Pset1}(c,d) and have only been able to achieve mitigation so far ($R_0^{\rm target} >1, \mu >0$). Data from \cite{data1} and the end date is June 10.}
\label{examples}
\end{figure*}

\subsection{Observation of strong and weak intervention in COVID-19 data}
 In Figs.~(\ref{examples}) we give some examples of data for number of new cases for nine countries where we see that some of the qualitative features seen in the model results in Fig.~\ref{Pset1}(a,b). In particular we see the fast exponential growth phase and then a much slower decay phase for the first six countries which have succeeded in controlling the disease with various levels of success. On the other hand we see that India, Brazil and Pakistan continue to show a positive $\mu$ and it is clear that intervention schemes need to be strengthened.

\begin{table*}[t]
\caption{Growth rate $\mu=0.05$: Predictions for India with different choices of parameter values.}
\begin{center}
\begin{tabular}{|c|c|c|c|c|c|c| c|c|}
\hline 
\hline
$\sigma$& $\gamma$ & $\alpha$ & $R_0$ (free) &$R^{\rm target}_0$  & Peak daily cases  (PDC) &  Time of peak & Total affected & Total deaths \\
\hline 
0.5 & 0.2&  0.67 & 2.28& 1.33  & 2,456,630  &  158 (2nd week September) & 45 \%  & 1,936,000\\
\hline
0.5& 0.2&  0.9& 2.16 & 1.3  & 686,770 &  132 (3rd week August) & 42 \%& 550,700 \\
\hline 
0.4 & 0.143 &  0.67 & 2.82 & 1.45  & 2,956,600  &  161 (3rd week September) & 55 \%  & 2,363,700\\
\hline
0.4 & 0.143 &  0.9& 2.65  & 1.41  & 832,890. &  136 (4th week August) & 52 \%& 676,140 \\
\hline
\end{tabular}
\end{center}
\label{table1}
\caption{Growth rate $\mu=0.05$: Predictions for Delhi with different choices of the asymptomatic fraction $\alpha$.}
\begin{center}
\begin{tabular}{|c|c|c|c|c|c|c|c|c|}
\hline 
\hline
$\sigma$& $\gamma$ & $\alpha$ & $R_0$ (free)& $R_0^{\rm target}$   & Peak daily cases  (PDC) &  Time of peak & Total affected & Total deaths \\
\hline 
0.5 & 0.2& 0.67  & 2.28& 1.33 &35,904 & 114 (1st week August) & 45 \%& 28,296 \\
\hline 
0.5& 0.2&  0.9 & 2.16 & 1.3 & 10,037 &  89 (2nd week July) & 42.3 \%& 8,048 \\
\hline 
0.4 & 0.143 & 0.67  & 2.82& 1.45 &43,212 & 118 (2nd week August) & 55 \% & 34,546 \\
\hline 
0.4& 0.143 &  0.9 & 2.65 & 1.41 & 12,173 & 93 (2nd week July) & 52 \%& 9,882 \\
\hline 
\end{tabular}
\end{center}
\label{table2}
\caption{Growth rate $\mu=0.05$: Predictions for Mumbai with different choices of the asymptomatic fraction $\alpha$.}
\begin{center}
\begin{tabular}{|c|c|c|c| c| c| c| c|c|}
\hline 
\hline
 $\sigma$& $\gamma$ & $\alpha$ & $R_0$ (free)& $R_0^{\rm target}$  & Peak daily cases  (PDC) &  Time of peak & Total affected & Total deaths \\
\hline 
0.5&0.2 & 0.67  & 2.28 & 1.33 &24,566 &  97 (3rd week July) & 45 \%& 19,360 \\
\hline 
0.5& 0.2 & 0.9 & 2.16 & 1.3 & 6,867 &  71 (4th week June) & 42.5 \%& 5,507 \\
\hline 
0.4 & 0.143 & 0.67  & 2.82 & 1.45 & 29,566 & 100 (3rd week July) & 55 \% & 23,637 \\
\hline 
0.4& 0.143 &  0.9 & 2.65 & 1.41 & 8,328 &  75 (4th week June) & 52 \%& 6,761 \\
\hline 
\end{tabular}
\end{center}
\label{table3}
\end{table*}

\begin{table*}[t]
\caption{Growth rate $\mu=0.035$: Growth rate $\mu=0.035$: Predictions for India with different choices of parameter values.}
\begin{center}
\begin{tabular}{|c|c|c|c|c|c|c| c|c|}
\hline 
\hline
$\sigma$& $\gamma$ & $\alpha$ & $R_0$ (free) &$R^{\rm target}_0$  & Peak daily cases  (PDC) &  Time of peak & Total affected & Total deaths \\
\hline 
0.5 & 0.2&  0.67 & 2.28& 1.23  & 1,331,400  &  208 (1st week November) & 34.5 \%  & 1,482,000\\
\hline
0.5& 0.2&  0.9& 2.16 & 1.207  & 368,809 & 172 (1st  week October) & 32 \%& 418,500 \\
\hline 
0.4 & 0.143 &  0.67 & 2.82 & 1.31  & 1,651,200  &  214 (1st week November) & 43 \%  & 1,856,300\\
\hline
0.4 & 0.143 &  0.9& 2.65  & 1.28  & 459,776 &  178 (1st week October) & 40.5 \%& 526,222 \\
\hline
\end{tabular}
\end{center}
\label{table4}
\caption{Growth rate $\mu=0.035$: Predictions for Delhi with different choices of the asymptomatic fraction $\alpha$.}
\begin{center}
\begin{tabular}{|c|c|c|c|c|c|c|c|c|}
\hline 
\hline
$\sigma$& $\gamma$ & $\alpha$ & $R_0$ (free)& $R_0^{\rm target}$   & Peak daily cases  (PDC) &  Time of peak & Total affected & Total deaths \\
\hline 
0.5 & 0.2& 0.67  & 2.28& 1.227 &19,458 & 146 (1st week September)  & 34.5 \%& 21,661 \\
\hline 
0.5& 0.2&  0.9 & 2.16 & 1.207 & 5,390 &  109 (4th week July) & 32 \%& 6,117 \\
\hline 
0.4 & 0.143 & 0.67  & 2.82& 1.31 &24,132 & 152 (1st week September) & 43 \% & 27,131 \\
\hline 
0.4& 0.143 &  0.9 & 2.65 & 1.28 & 6,719 & 116 (1st week August) & 40.5 \%& 7,691 \\
\hline 
\end{tabular}
\end{center}
\label{table5}
\caption{Growth rate $\mu=0.035$: Predictions for Mumbai with different choices of the asymptomatic fraction $\alpha$.}
\begin{center}
\begin{tabular}{|c|c|c|c| c| c| c| c|c|}
\hline 
\hline
 $\sigma$& $\gamma$ & $\alpha$ & $R_0$ (free)& $R_0^{\rm target}$  & Peak daily cases  (PDC) &  Time of peak & Total affected & Total deaths \\
\hline 
0.5&0.2 & 0.67  & 2.28 & 1.23 &13,316 & 120  (1st week August) & 34.5 \%& 14,821 \\
\hline 
0.5& 0.2 & 0.9 & 2.16 & 1.21 & 3,688 &  84 (1st week July) & 32 \%& 4,185 \\
\hline 
0.4 & 0.143 & 0.67  & 2.82 & 1.31 & 16,512 & 126 (2nd week August) & 43 \% & 18,563 \\
\hline 
0.4& 0.143 &  0.9 & 2.65 & 1.28 & 4,597 & 90 (2nd week July) & 40.5 \%& 5,262 \\
\hline 
\end{tabular}
\end{center}
\label{table6}
\end{table*}

\section{Difficulties in making predictions from the extended SEIR model: case study for India}
\label{sec:india}
 In the following we make some heuristic predictions, based  on the analytic results in Eqs.~(\ref{r0eq},\ref{eq:peakI},\ref{eq:peakT}) and the present observed data, for daily new cases in India ($N \approx 1.3\times 10^9$), in the state of Delhi ($N\approx 1.9 \times 10^7$) and  in the city of Mumbai ($N \approx 1.3\times 10^7$).  The analysis here is based on the assumption of a best case scenario where the value of $R_0^{\rm target}$, achieved after various intervention schemes is maintained at a constant value.

Here we assume that intervention has effectively been through SD, with $r << \gamma$ being neglected.  We consider the following choice of  parameter  values which appears to be reasonable for getting a conservative estimate: $\sigma =1/2, \tilde{\beta}_p=\beta, \tb_a=2\beta/3, \tg_p=\gamma_p=\gamma, \tg_a=\gamma_a=3 \gamma/2 $, i.e, we assume that asymptomatics are $2/3$rd less infectious and recover $3/2$ times faster. This gives us [using Eq.~\eqref{gammaeff}]  $\gamma_e = \gamma/(1-\alpha/3)$ and  the effective reproductive number as $R_0^{\rm target}=(1-5\alpha/9) \beta/\gamma$. From this last relation we can write $\beta= \gamma R_0^{\rm target}/(1-5 \alpha/9)$. Plugging this into the equation for the eigenvalues, Eq.~\eqref{roots}, 
and replacing $\lambda$ by the  observed  mean exponential growth rate (we choose the values of $\mu= 0.05$ and  $\mu =0.035$ which are representative of the values observed in India since around April 10), we see that we basically get an equation for $R^{\rm target}_0$ in terms of $\alpha, \sigma, \gamma$ and $\mu$. For specific choices of $\alpha, \sigma$ and $\gamma$, the observed values of $\mu$ before and after intervention will then give us the corresponding values of $R_0$.

For our analysis we need to know the total infections $I(0)$ on some day (we take this as April 11) and we estimate it in the following way. Suppose that the daily observed cases on this day was $F_p(0)$ (assuming that only the symptomatics are detected). Then we have $I_p(0)=F_p(0)/\gamma_p$. From Eq.~\eqref{I_ap^max} we have $I_p^{(m)}=(1-\alpha)\gamma_e I^{(m)}/\gamma_p$ and so the time to the peak can be estimated as $t^{(m)}=\mu^{-1}\ln [I_p^{(m)}/I_p(0)]$.
We use Eq.~(\ref{eq:peakI}) to compute the peak number of infections $I^{(m)}$ and the peak daily cases (PDC) is then obtained as PDC$=F_p^{(m)}=\gamma_p I_p^{(m)}=(1-\alpha) \times \gamma_e \times I^{(m)}$.   The total affected population fraction $\bar{x}$, can be computed from Eq.~\eqref{r0eq}, using only the knowledge of $R_0^{\rm target}$.  If we assume the number of deaths is $1\%$ of all symptomatic cases this gives us an estimate for the total number of deaths  as  $N \bar{x} (1-\alpha)/100$.  

\begin{figure}[t]
\center
\includegraphics[scale=0.3]{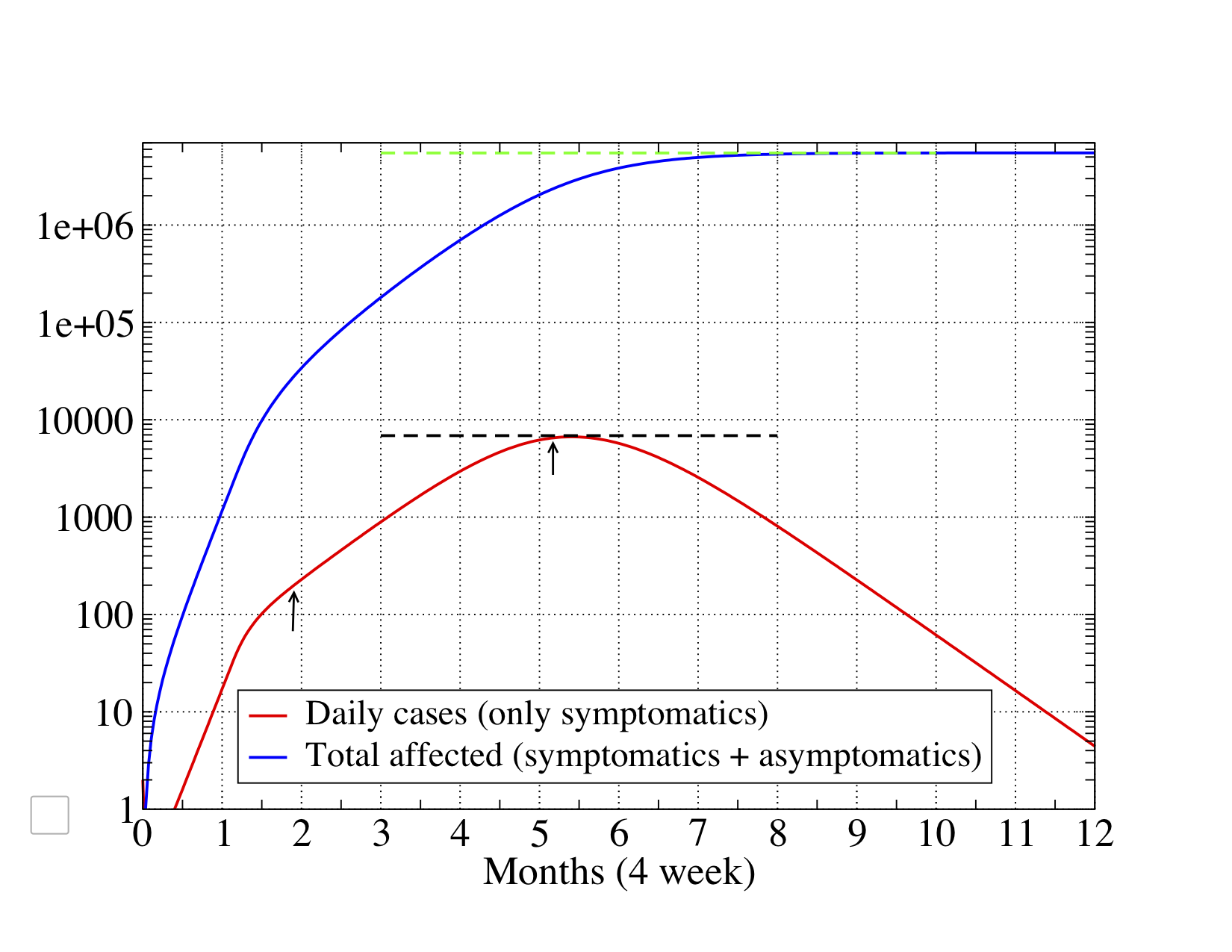}
\caption{ Plot of the new daily cases (${\gamma}_p I_p$)  and total affected population fraction ($R$), as a function number of months for one of the parameter sets in Table.~\ref{table3} for the city of Mumbai. Parameter values were $\sigma=0.5, \tilde{\gamma}_p=0.2, \alpha=0.9$, $R_0=2.16$ before intervention and $R_0^{\rm target}=1.3$. The dashed lines give the analytic predictions for the peak daily cases (black line) and the final affected population ( green line), and show the good agreement with the numerics. The arrows indicate the date when initial condition was specified $\gamma_p I_p(0)=195$ and the peak infection date, which occurs about 15 days after the date predicted from Eq.~\eqref{eq:peakT}. }
\label{fig:india}
\end{figure}

The observed daily new cases in India, Delhi and Mumbai on April 10 were around $F_p(0)\approx 900$, $F_p(0) \approx 115$ and $F_p(0)\approx 195$ respectively \cite{indiandata}.  For a range of choice of the parameters with $\sigma=0.5,~0.4$,   $\gamma=0.2,~0.143$ and of $\alpha = 0.67,~0.9$, and two representative values of the post-intervention growth rates, $\mu=0.05, 0.035$, we compute the corresponding values obtained for $R_0$ and $R_0^{\rm target}$. These and the estimates  for  PDC$=F_p^{(m)}, t^{(m)}$ and $\bar{x}$ are given in Tables~(\ref{table1}~\ref{table3}) for $\mu=0.05$ and in Tables~(\ref{table4}-\ref{table6}) for $\mu=0.035$, for India, Delhi and Mumbai. Note that while the peak numbers and total affected population and deaths simply scale with population size, the time to peak depends on the daily detected numbers on April 10, and this leads to the observed differences in the time to the peak for the three cases. We also note here that changing the initial conditions by about $10\%$ causes a change of few days in the peak time while the other quantities remain unchanged.
The full numerical solution in Fig.~\eqref{fig:india} also shows that the complete suppression of the disease takes more than 6 months after the peak.

In Fig.~\ref{fig:india} we show results of a numerical solution of the dynamical equations in presence of intervention (SD) for one of the parameter sets in Table.~\ref{table3} and find excellent agreement with our analytic formula in Eqs.~(\ref{r0eq},\ref{eq:peakI}-\ref{gammaeff}). We see that the predicted peak time is off by about $10\%$. The numerics also shows that the complete suppression of the disease takes more than 6 months after the peak.

We point out that the mixed-population assumption of the SEIR model is expected to be more accurate for a smaller population and so the estimates for Delhi and Mumbai would be more reliable than the one for India. For a big and highly in-homogeneous country like India, smaller regions (states or cities)  would have  different values of $\mu$ and $R_0$ and also different initial conditions, hence the global values would not capture the local dynamics correctly. It is likely that the numbers in Table~\ref{table1}   are an over-estimate of the true future trajectory. For the state of Delhi and the city of Mumbai these would be more accurate, \emph{however we see that the uncertainty in the true value of $\alpha$ and other parameters leads to a huge uncertainty in the predictions}.
\vspace{1em}

\section{Discussion}
\label{sec:conclusions}

 Several earlier work have discussed, using determinsitic compartmentalized models,  the effect of asymptomatic affected population and the effect of intervention measures on the COVID pandemic~\cite{Tang2020,colaneri2020,gatto2020,india1,india2,india3,india4,india5,frank2020a,piovella2020,israel2020,nigel2020}.   Here we  present a somewhat different choice of compartments and perform a careful quantitative comparison of different intervention strategies. A distinction from earlier studies is that we present a number of analytic results which we believe will be useful beyond their immediate application to epidemiological modeling, in more general studies of population dynamics.

To  summarize our findings, a modified version of the SEIR model, incorporating asymptomatic individuals, was analyzed in detail. We have obtained a number of analytical results for the full nonlinear model which can be useful in making empirical estimates of various  important quantities  that provide information on  disease progression.  We believe that the derivation of these results can be extended to more sophisticated SIR type models including more compartments and more complex interactions. From the linearized dynamics we point out a simple but important property, namely that at early times the motion of the system quickly settles along the direction of the dominant eigenvector. This allows one to determine accurately initial conditions from sparse data. 
 We  provided numerical examples to illustrate these  ideas and in addition have  provided comparisons with real COVID-19 data. Looking at COVID-19 data in several countries, we find that the  extended SEIR model captures some important qualitative features and hence could provide  guidance in policy-making

We used the extended SEIR  model for analyzing the effectiveness of different intervention protocols in controlling the growth of the COVID-19 pandemic. Non-clinical interventions can be either through social distancing or testing-quarantining. Our results indicate that a combination of both, implemented over an extended  period may be the most effective and practical strategy.  
We have attempted to relate real testing rates to the parameters of the model and comment on what  the minimum testing rates should be in order for testing-quarantining to  be an effective control strategy.

Finally we have used our analytic formulas to make predictions for disease peak numbers and expected time to peak for India, the state of Delhi and the city of Mumbai, pointing out that these predictions would be highly unreliable for India (due to big inhomogeneity in disease progression across the country) and perhaps more reliable for the cases of Delhi and Mumbai. Our main conclusion here is that the lack of precise knowledge of the disease  parameters (e.g the fraction of asymptomatic carriers)  and changing control strategies lead to rather large uncertainties in the predictions. Nevertheless, we believe that they could perhaps be used to obtain reasonable bounds.

\begin{acknowledgments}
We thank Jitendra Kethepalli and Kanaya Malakar for very helpful discussions and Ranjini Bandyopadhyay, Siddhartha Chatterjee, Joel Lebowitz, Srujana Merugu, Alpan Raval, Sankar Das Sarma and Sriram Shastry  for a careful reading of the draft and making useful suggestions. We acknowledge support of the Department of Atomic Energy, Government of India, under project no.12-R$\&$D-TFR-5.10-1100.
\end{acknowledgments}

\end{document}